\documentclass[conference]{IEEEtran}
\usepackage{cite}
\usepackage{amsmath,amssymb,amsfonts}
\usepackage{algorithmic}
\usepackage{graphicx}
\usepackage{textcomp}
\usepackage{xcolor}
\def\BibTeX{{\rm B\kern-.05em{\sc i\kern-.025em b}\kern-.08em
    T\kern-.1667em\lower.7ex\hbox{E}\kern-.125emX}}

\usepackage[capitalize]{cleveref}
\usepackage{booktabs}
\usepackage{listings}
\usepackage[inline]{enumitem}

\usepackage[newfloat, frozencache]{minted}
\newlength{\mintednumbersep}
\AtBeginDocument{%
  \sbox0{\tiny00}%
  \setlength\mintednumbersep{\parindent}%
  \addtolength\mintednumbersep{-\wd0}%
}
\newminted[mlir]{CustomLexers.py:MLIRLexer -x}{mathescape,fontsize=\scriptsize,linenos,xleftmargin=\parindent,numbersep=\mintednumbersep,frame=single,rulecolor=\color{gray!40}}
\newminted[sycl]{CustomLexers.py:SYCLLexer -x}{mathescape,fontsize=\scriptsize,linenos,xleftmargin=\parindent,numbersep=\mintednumbersep,frame=single,rulecolor=\color{gray!40}}

\newcommand\blfootnote[1]{%
  \begingroup
  \renewcommand\thefootnote{}\footnote{#1}%
  \addtocounter{footnote}{-1}%
  \endgroup
}

\usepackage{microtype}

\newcommand{\simpleauthor}[4]{
\IEEEauthorblockN{#1}
\IEEEauthorblockA{\textit{#2} \\
#3 \\
#4}
}

\begin{document}

\title{Experiences Building an MLIR-based SYCL Compiler}

\author{
\simpleauthor{Ettore Tiotto}{Intel Corporation}{Toronto, Canada}{ettore.tiotto@intel.com}
\and
\simpleauthor{Víctor Pérez}{Codeplay Software}{Edinburgh, Scotland, UK}{victor.perez@codeplay.com}
\and
\simpleauthor{Whitney Tsang}{Intel Corporation}{Toronto, Canada}{whitney.tsang@intel.com}
\and
\simpleauthor{Lukas Sommer}{Codeplay Software}{Edinburgh, Scotland, UK}{lukas.sommer@codeplay.com}
\and
\simpleauthor{Julian Oppermann}{Codeplay Software}{Edinburgh, Scotland, UK}{julian.oppermann@codeplay.com}
\and
\simpleauthor{Victor Lomüller}{Codeplay Software}{Edinburgh, Scotland, UK}{victor@codeplay.com}
\and
\simpleauthor{Mehdi Goli}{Codeplay Software}{Edinburgh, Scotland, UK}{mehdi.goli@codeplay.com}
\and
\simpleauthor{James Brodman}{Intel Corporation}{Boston, MA, USA}{james.brodman@intel.com}
}

\maketitle

\begin{abstract}
Similar to other programming models, compilers for SYCL, the open programming model for heterogeneous computing based on C++, would benefit from access to higher-level intermediate representations. The loss of high-level structure and semantics caused by premature lowering to low-level intermediate representations and the inability to reason about host and device code simultaneously present major challenges for SYCL compilers. 

The MLIR compiler framework, through its dialect mechanism, allows to model domain-specific, high-level intermediate representations and provides the necessary facilities to address these challenges.

This work therefore describes practical experience with the design and implementation of an MLIR-based SYCL compiler. By modeling key elements of the SYCL programming model in host and device code in the MLIR dialect framework, the presented approach enables the implementation of powerful device code optimizations as well as analyses across host and device code.

Compared to two LLVM-based SYCL implementations, this yields speedups of up to 4.3x on a collection of SYCL benchmark applications.

Finally, this work also discusses challenges encountered in the design and implementation and how these could be addressed in the future.
\end{abstract}


\section{Introduction}\label{sec:intro}
For many applications in the machine learning and high-performance computing (HPC) domain, heterogeneous computing has become inevitable to meet the computational demands, creating a strong need for powerful heterogeneous programming models. 

There are several heterogeneous programming models, such as CUDA, SYCL, and OpenCL, enabling offloading portions of an application on hardware accelerators. Among them, SYCL provides a modern C++-based open-standard single-source programming paradigm enabling portability across a wide range of hardware from different vendors. SYCL supports kernel dependency tracking and data management between host and device that helps reduce boilerplate code found in many similar programming models, e.g., OpenCL.

At the same time, being based on C++ also means that SYCL compilers are presented with the same challenges as other C++-based languages. One such challenge is to keep track of the high-level structure of the program and domain-specific information that can significantly affect the effectiveness of compiler optimizations. Hence, the SYCL semantics, including work-item parallel execution and device memory access, will be lost in the early stages of the compilation pipeline when lowering to a low-level intermediate representation (IR) such as LLVM IR.

Also, in most current SYCL compilers, translation and optimizations of the device code happens in isolation from host compilation. Such separation can prevent passing relevant information, e.g., the invocation context of a device kernel, to the device optimization pipeline.

The MLIR compiler framework is well suited to address these challenges. MLIR dialects allow capturing domain-specific semantics, in the case of this work the SYCL semantics, on a high level of abstraction. At the same time, MLIR's ability to nest operations allows reasoning about host and device code at the same time.

Building on top of the MLIR framework~\cite{mlir}, this work therefore aims to address these challenges. By leveraging MLIR's more fine-grained lowering process and extensibility, which allows capturing the semantics of the SYCL programming model and making it accessible to compiler optimizations, the SYCL-MLIR project presented in this work aims to build more powerful optimizations for SYCL code.

Overall, this work makes the following contributions:
\begin{itemize}
    \item Extension of the MLIR framework to capture the semantics of the SYCL programming model.
    \item Architecture and implementation of an MLIR-based compilation flow that allows joint analysis of host and device code to enable better device optimizations based on invocation context.
    \item Design and implementation of analyses for SYCL host and device code as well as device code optimizations.
    \item Evaluation of the MLIR-based compilation flow with a variety of SYCL benchmark applications and comparison with an existing LLVM-based SYCL compiler.
    \item Report on our practical experience with building an MLIR-based compiler for a C++-based parallel programming model and the main challenges involved.
\end{itemize}
The rest of this work is structured as follows. \cref{sec:background} introduces necessary background information on the SYCL programming model and the MLIR compiler framework. \crefrange{sec:aim}{sec:raise-host} describe the SYCL-MLIR project, its extension of MLIR for SYCL, the compilation flow as well as the host and device analyses and optimizations in detail. \cref{sec:eval} evaluates the MLIR-based SYCL compiler and compares it to an existing LLVM-based SYCL compiler. Related approaches in literature are discussed in \cref{sec:related}, and \cref{sec:conclusion} concludes this work and gives an outlook on future development.

\section{Background}\label{sec:background}
\subsection{SYCL}\label{sec:sycl}

SYCL~\cite{sycl} is a Khronos open standard defining a C++-based parallel programming model for heterogeneous systems, providing a rich API
for users to access accelerator resources.

While SYCL shares some of the underlying concepts and abstractions with them, there are a number of interesting differences between SYCL and other heterogeneous, parallel programming models such as OpenCL, OpenMP or CUDA\@. In contrast to these programming models, SYCL does not rely on extensions to the C++ language.

SYCL kernels are expressed using functors (plain C++ class defining a call operator
or a lambda expression).
A kernel is submitted for execution to a \emph{queue} using a \emph{command-group function}
(shortened to command-group).
The command-group allows the user to express the dependencies
the kernel requires in order to be run. These dependencies can involve waiting on another kernel
or requesting a memory transfer.
The command-group will finish with the submission of the kernel specifying the index space it shall run on.

Kernels are submitted for execution with an index space (called ND-range) subdivided into \emph{work-groups}.
For each point of this space, an instance (called \emph{work-item}) executes the kernel
in parallel to other work-items.
Work-items are also bundled into work-groups, allowing some resource sharing,
communication and synchronization via barriers between work-items
belonging to the same work-group.
Each SYCL kernel functor will take as a parameter either an \texttt{id}, \texttt{item} or
\texttt{nd\_item} encapsulating the position of the work-item within the ND-range.

SYCL also defines a memory hierarchy with different levels of visibility: global, local
and private memory. Whereas global memory is shared by all work-items, local memory
is shared by all work-items \emph{within a work-group},
enabling work-group--wide cooperation, and it is usually faster to access than global memory.
Finally, private memory is only visible to a given work-item.

SYCL offers two ways to manage memory between host and device: Unified Shared Memory (USM) and the buffer and accessor model.
When using USM, the user manipulates pointers directly. The memory is allocated and freed using dedicated \texttt{malloc}- and \texttt{free}-like functions
and the user is responsible for transferring data to and from the device manually.
The buffer and accessor model allows the SYCL runtime to handle not only memory management across devices
(memory creation, deletion, transfer to and from device) to ensure memory consistency but
also the creation of dependencies between kernels.

A SYCL buffer is a multi-dimensional container owning the memory that tracks where copies live across the host and devices,
however it does not provide access to the memory directly. In order to get access to the data,
the user must create an accessor object inside a command-group.
The creation of the object will create a dependency within the SYCL scheduler in order to ensure the data is available prior to the kernel execution.
Inside the kernel, the user can use the accessor object in a similar way to a regular C++ vector or \emph{mdspan} container.

The accessor is a heavy object encapsulating several dynamic pieces of information: the pointer to the data, the full range of the data but also a sub-range and offset.
The sub-range and offset is only useful in the case of a \emph{ranged accessor} which allows the user to pass only part of a buffer to the kernel.
However, the distinction between a ranged and non-ranged accessor is done by calling different constructors and  is never reflected on the C++ type.
Other information is also embedded statically via template parameters, including if it is read-only, write-only or read-write.

While the underlying concepts such as the execution and memory model are similar to OpenCL or CUDA, the buffer and accessor model is a good example of how SYCL leverages the power of modern C++ to automate some of the cumbersome tasks in heterogeneous programming. If an application uses the buffer and accessor abstraction, the SYCL runtime can fully automate dependency tracking between kernels and necessary data movements between host and one or multiple devices, whereas in more low-level programming models such as OpenCL or CUDA, the developer needs to manually take care of these details. This example, and other more high-level abstractions in SYCL such as the powerful out-of-order queues, make typical SYCL more concise, improving maintainability and programmer productivity.

The higher level of abstraction can also facilitate portability between different architectures. In contrast to proprietary models such as CUDA, which can only run on devices from one specific vendor, SYCL's open nature allows implementations to target hardware from different vendors, making SYCL available on many different platforms.

While a comparison of SYCL with other programming models in even more detail is beyond the scope of this work, such comparisons can for example be found in~\cite{compare1,compare2,compare3,compare4,compare5,compare6}.

\subsection{MLIR}\label{sec:mlir}
The inception of the MLIR framework~\cite{mlir} was motivated by the insights that prematurely lowering a high-level programming model to a low-level representation can significantly harm the compiler's ability to perform powerful optimizations, and that many of the existing IR representations (e.g., LLVM IR) are hard to extend. Instead of working on a low-level representation such as LLVM IR, many optimization passes would benefit from a higher-level representation that preserves the structure of the program, e.g., loops, and can also encode domain knowledge about the application. 

This insight is also an important motivation for this work, which seeks to improve SYCL code optimization through an intermediate representation that allows encoding SYCL's semantics directly in it.

Even before MLIR, the benefit of having several intermediate representations in a compiler infrastructure was well recognized in the literature~\cite{tobey,Novillo2004DesignAI,Merrill2003GenericAG}. Programming languages such as Swift or Rust later introduced higher-level representations with the same aim, before eventually lowering it to LLVM IR\@. MLIR now seeks to provide a platform and common infrastructure for such representations, to enable better re-usability across different frontends.

The core concept that allows MLIR to cover such a wide range of abstraction levels and domains is the notion of \emph{dialects}. A dialect encapsulates the \emph{attributes}, \emph{types} and \emph{operations} associated with the representation of a specific domain. The upstream MLIR project collects a number of composable dialects that are useful across different abstractions, such as the \texttt{arith} dialect for arithmetic operations or the structured control flow (\texttt{scf}) dialect for loops. In addition to the upstream dialects, users of MLIR can freely extend the framework by defining specialized dialects to fit their problem domain. To this end, this work defines a dialect capturing SYCL semantics, with more details given in \cref{sec:flow}.

MLIR-based compilation flows will then typically use a combination of multiple dialects to represent an application at every stage of the compilation process. Once optimizations on one level of abstraction are completed, the representation is typically \emph{lowered} to another set of dialects, where more optimizations can happen. Overall, this yields a gradual lowering process through dialect conversion and pattern rewriting, this aspect is also exercised in this work as described in \cref{sec:flow}.

In addition to dialect abstraction and progressive lowering, MLIR also provides a common infrastructure for creating analyses and transformation passes. This common infrastructure also underpins the analyses and transformations for this work, described in \crefrange{sec:analyses}{sec:raise-host}.

\section{Aim \& Approach}\label{sec:aim}
As already briefly discussed in \cref{sec:intro}, the aim of the SYCL-MLIR project presented in this work is to build an MLIR-based compiler for the SYCL heterogeneous programming model. There are two main motivations for the use of the MLIR compiler framework for a SYCL compiler. 

First, the ability to nest operations that is inherent to MLIR allows the compilation flow to represent SYCL host and device code alongside each other in the same MLIR module, enabling better analyses of device kernels in the context of their invocation on the host. The details of the compilation flow and how this side-by-side analysis of host and device code is achieved are described in \cref{sec:flow,sec:raise-host}.

Second, the MLIR concept of dialects enables the compilation flow to initially preserve the high-level semantics of the SYCL programming model. This way, the precise semantics of the language can be made available to analyses and transformations, and lowered only after optimizations benefiting from access to the SYCL semantics have concluded. To this end, the SYCL-MLIR project defines a SYCL dialect, modeling key entities of the SYCL programming model as MLIR attributes, types and operations. 

In device code, the SYCL dialect models two main concepts. One is the position of the current work-item (see \cref{sec:sycl}) in the overall execution grid and within its work-group. To this end, the SYCL classes \texttt{id}, \texttt{item}, \texttt{nd\_item}, \texttt{range}, \texttt{nd\_range} and \texttt{group} are modeled as types in the SYCL dialect and key functions to obtain and operate on instances of these classes are represented as operations of the dialect. 

Another important part of the SYCL programming model on devices with high relevance for compiler transformations is the access to device memory through SYCL accessors (see \cref{sec:sycl}). To enable optimizations to reason about and transform memory access behavior of a kernel, the SYCL \texttt{accessor} class is modeled as another type in the SYCL dialect and operations such as accessing an element of memory through an accessor are modeled as MLIR operations in the dialect. Typically, memory is accessed based on the position of the work-item in the execution grid, so the representation of \texttt{id} and related classes in the SYCL dialect play another important role here.

On the host side, the dialect aims to represent the invocation context of kernels, including the arguments passed to those kernels and the ND-range. To this end, operations representing the construction of the SYCL command-group and the kernel function object (see \cref{sec:sycl}) are added to the dialect. Many of the types introduced for the representation of SYCL entities on the device side, e.g., for the \texttt{id} or \texttt{range} classes, can be reused here, and additional types for classes such as \texttt{buffer} have also been added to the dialect. 

Next to the kernel launch and its arguments, the provenance of accessors used in device code can provide valuable information. Therefore, operations modeling the construction of accessors and their underlying buffers (see \cref{sec:sycl}) are also added to the dialect. Analyzing the parameters used for construction of accessors and buffers can provide useful insights into their behavior, for example into potential overlap or aliasing of two accessors. 

The raising process for the host code and how the aforementioned operations are used there is described in greater detail in \cref{sec:raise-host}.

\section{Compilation flow}\label{sec:flow}

\begin{figure}[t]
\centering
\includegraphics[width=.83\columnwidth]{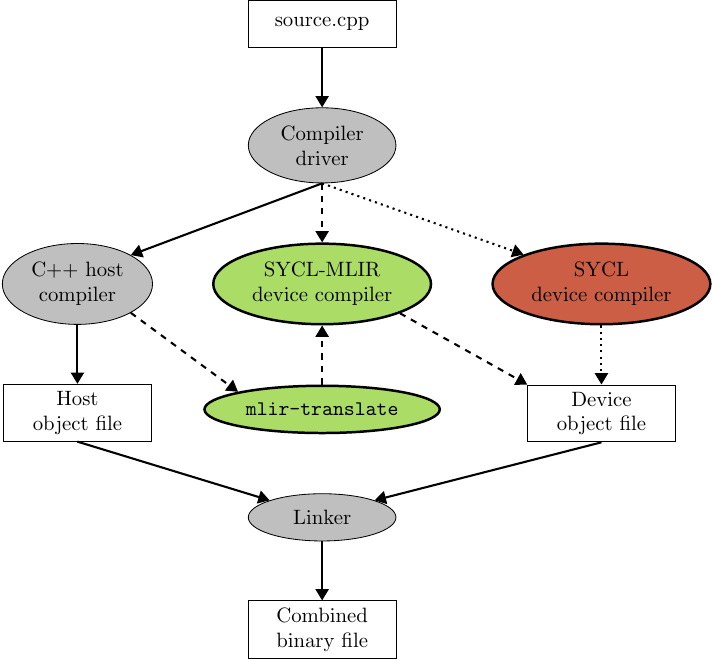}
\caption{SYCL compilation flow. Solid lines and elements in white and gray show common parts of the compilation flow shared by DPC++ and SYCL-MLIR\@. DPC++'s (dotted path) device compiler only receives the kernel source code as an input and device compilation happens entirely in isolation from host compilation. SYCL-MLIR (dashed path), on the other hand, receives an MLIR representation of the host module obtained via \texttt{mlir-translate}, enabling joint analysis of host and device.}
\label{img:combined-comp}
\end{figure}

The SYCL standard (see \cref{sec:sycl}) describes a \emph{single-source multiple compiler passes} (SMCP) compilation flow. This consists in
processing the same source code several times using different compilers: one for the host, and one for each target. In the case of the Intel DPC++ compiler~\cite{dpcpp}, the generated device images are linked together and wrapped in the host object to be later read by the SYCL runtime (dotted path in \cref{img:combined-comp}).

However, this SMCP technique presents a limitation by design: no host-device cross-boundary optimizations can be efficiently performed at compile time, due to the limited information being shared between host and device compilers. To address this point, using a custom SYCL compiler handling both host and device code, i.e., through the \emph{single-source single compiler pass} (SSCP) approach, opens up the possibility to perform transformations in the host code using device code analysis, and vice versa.

To translate from C++ AST to MLIR we use our own Polygeist~\cite{polygeist} fork as a device compiler.
While being efficient for compiling the device side, it struggles to handle the required C++ construct of the host side. For example, virtual functions and exceptions are not properly supported.
In order to obtain similar results to the SSCP technique without needing a robust C++ MLIR frontend (see \cref{sec:related}), we take an alternative approach obtaining MLIR host code from LLVM IR, which can then be used to leverage device compilation.
We then obtain a joint representation of host and device code exploiting MLIR's nested IR structure (see dashed path in \cref{img:combined-comp}). This way, we set the basis for a SYCL compiler capable of performing device code optimizations using host code analysis.

After this, a series of custom analyses and transformation passes using SYCL domain-specific knowledge are executed, including a 
\emph{host raising pass} detecting and raising relevant patterns in SYCL host code, as described in \cref{sec:raise-host}. This alternative approach enables host-device optimizations at little cost compared to the development of a C++ MLIR frontend. However, as a downside, changes to SYCL runtime code can lead to raising pattern matching to fail, forcing this pass to be up-to-date with runtime changes.

\section{Compiler Analyses}\label{sec:analyses}
During the development of the SYCL-MLIR compiler we have found the need to extend the MLIR framework with several static analyses and utilities that are generally useful in an optimizing compiler. This section gives a brief overview of our implementation, highlighting how the semantic information encoded by several SYCL dialect operations have been leveraged.

\subsection{Alias Analysis}\label{sec:alias-analysis}

MLIR provides an alias analysis framework which can be readily augmented to take into account the domain knowledge provided by different dialects. In order to leverage the semantic information of the SYCL dialect, we have created a SYCL-specific alias analysis extending MLIR's existing one.

The semantics of SYCL dialect operations can be encoded in the analysis, allowing the compiler to prove that values yielded by SYCL operations do not alias in many circumstances, and thus leveraging the SYCL dialect to make alias analysis more precise.

\subsection{Reaching Definition Analysis}\label{sec:reach-def-anal}

The classic reaching definition analysis is a data-flow analysis designed to provide the set of operations that might have modified a value in memory at a given program point. Our implementation leverages the data-flow analysis framework provided by MLIR, and our specialized alias analysis. We consider two kinds of reaching definitions for a value:
  \begin{itemize}
    \item \emph{modifiers (MODS)}: definitions for the value itself or definition of a value that is known to be definitely (must) aliased to it
    \item \emph{potential modifiers (PMODS)}: definitions of a value that is known to be possibly (may) aliased to the value
  \end{itemize}

\begin{listing}[t]
\begin{mlir}
func.func @foo(
  scf.if 
    memref.store 
  } else {
    memref.store 
  }
  ... = memref.load 
}
\end{mlir}
\caption{Function with potentially aliasing \texttt{memref} arguments.}\label{lst:reaching-definition-example}
\end{listing}

As an example let us consider the MLIR code snippet in \cref{lst:reaching-definition-example}.
The store operation at line 4 updates the memory location \texttt{\%ptr1} points to directly, while the store operation at line 6 might update the same memory location because \texttt{\%ptr2} may alias \texttt{\%ptr1}. It follows that the reaching definition for \texttt{\%ptr1} at line 8 is \texttt{\{MODS: a, PMODS: b\}}.

To define the memory effects of different operations, the MLIR framework provides a generic interface for operations, so analyses such as the reaching definition analysis can reason about effects of operations from different dialects.

By implementing this interface for the relevant SYCL dialect operations, the reaching definition analysis can be customized to account for the precise memory semantics of each SYCL dialect operation. 

\subsection{Uniformity Analysis}\label{sec:uniformity-analysis}

In GPU programming, a divergent branch is a branch in which the condition does not yield the same result for all work-items in the work-group (cf. \cref{sec:sycl}), causing subsets of work-items to branch to different program points.

Divergent branches can be recognized by tracking the \emph{uniformity} of variables. A variable is said to be \emph{uniform} if all work-items in a work-group assign the same value to it, and \emph{non-uniform} otherwise. A trivial example of a non-uniform value is the result of an operation yielding the global id of the work-item. Values assigned to the same memory location under divergent paths yield data divergence and may cause divergent control flow if used in a conditional expression. 

In the example in \cref{lst:uniformity-example}, the \texttt{\%gid\_x} value (line 9) is a source of non-uniformity, as it is the result of evaluating the \texttt{sycl.nd\_item.get\_global\_id} operation, which yields the global id (see \cref{sec:sycl}) of a SYCL work-item. As a result, the branch conditions \texttt{\%cond} (line 12) and \texttt{\%cond1} are therefore non-uniform, the former because it uses the non-uniform value and the latter because it loads from  the memory pointed to by \texttt{\%alloca}, which is assigned different values in the divergent branch (lines 14 and 16).

In order to reason about divergent control flow we have implemented the \emph{Uniformity Analysis} as an inter-procedural data-flow analysis based on MLIR's data-flow framework.

Formal parameters are initially assigned \emph{unknown} uniformity, with the exception of the SYCL kernel entry point which has parameters that are \emph{uniform} by definition. The analysis then propagates the uniformity of values by visiting operations in a function. A custom trait informs the analysis about SYCL operations that are known sources of non-uniformity.

The uniformity of values yielded by other operations is:
\begin{itemize}
    \item \emph{non-uniform}: if any operand is non-uniform;
    \item \emph{unknown}: if any operand has unknown uniformity, or
    \item \emph{uniform}: if all operands are uniform and the operation is free of memory effects.
\end{itemize}

Through the trait mechanism, the MLIR framework allows to leverage domain-specific knowledge to inform the analysis of the effects of dialect operations. The trait can easily be added to operations from other dialects to allow the uniformity analysis to also reason about operations from other dialects, demonstrating the re-usability facilitated by the MLIR framework.

Operations that have memory effects are further analyzed, using the memory effect interface that is also used for the reaching definition analysis. This way, the analysis cannot only reason about the SYCL dialect, but also all other dialect operations that implement this interface.

\begin{listing}
\begin{mlir}
func.func @non_uniform(
  // Get the global id of an nd_item (non-uniform value).
          : (memref<?x!sycl_nd_item_2>, i32) -> i64
  // The branch condition is non-uniform.
  scf.if 
    memref.store 
  } else {
    memref.store 
  }
  // Yields non-uniform value.
  // Divergent branch.
  scf.if 
    ...
\end{mlir}
\caption{Function showing a divergent branch.}\label{lst:uniformity-example}
\end{listing}

If the operation has unknown memory effects, it is conservatively considered to have \emph{unknown} uniformity. Otherwise, each memory effect is analyzed. For a write memory effect, using the \emph{Reaching Definition Analysis} (\cref{sec:reach-def-anal}), \emph{unknown} or \emph{non-uniform} uniformity is propagated from the (potential) modifiers and their dominating branch conditions.

Finally, the analysis works inter-procedurally by using the call graph to propagate the uniformity of the actual argument to a function across every possible call site. If all call sites are known (no external calls are possible), the uniformity of the parameters is computed by merging the uniformity of the actual arguments. 

In the SYCL-MLIR compiler, this analysis is currently used by the \emph{Loop Internalization} optimization (\cref{sec:loop-internalization}) to determine whether a loop is executed in a divergent region. This is a necessary prerequisite because the transformation needs to inject a group barrier, which would deadlock if it were to be executed in a divergent region.

\subsection{Memory Access Analysis}\label{sec:memory-access-analysis}
Reasoning about the memory access pattern of a kernel is key to many transformations, including some of the optimizations presented in \cref{sec:opt-device}. To this end, we have implemented a \emph{Memory Access Analysis} to derive the access pattern that SYCL memory accesses exhibit in a GPU kernel.

The analysis is based on~\cite{mem-access-pattern} and extends it to consider SYCL memory accesses. Given an affine loop, SYCL memory access patterns can be modeled by using an access matrix and a vector of offsets.

\begin{listing}[t]
\begin{mlir}
func.func @mem_acc(
      : (memref<?x!sycl_item_2>, i32) -> index
      : (memref<?x!sycl_item_2>, i32) -> index
  affine.for 
    // [gid_x+1, 2*i, 2*i+2+gid_y]
    sycl.constructor @id(
        : (memref<1x!sycl_id_3>, index, index, index)
        : (memref<?x!sycl_accessor_3_f32>, 
           memref<1x!sycl_id_3>) -> memref<?xf32>
  }
  return
 }
\end{mlir}
\caption{MLIR memory access example.}\label{lst:mem-access-example}
\end{listing}

As an example, consider the MLIR function in \cref{lst:mem-access-example}.
The memory access in the loop (line 23) uses a \texttt{memref} obtained via a \texttt{sycl.accessor.subscript} operation with an index given by the 3-dim SYCL id constructed at line 18 by a \texttt{sycl.constructor} operation. The indexing function is an affine function of the work-item global id  (\texttt{\%gid\_x} and \texttt{\%gid\_y}) and the loop induction variable \texttt{\%i} and, in the analysis, is described by an access matrix and an offset vector as follows:

$$
\begin{pmatrix}
    1 & 0 & 0 \\
    0 & 0 & 2 \\
    0 & 1 & 2
\end{pmatrix}
\times
\begin{pmatrix}
    \%gid\_x \\
    \%gid\_y \\
    \%i
\end{pmatrix}
+
\begin{pmatrix}
    1\\
    0 \\ 
    2
\end{pmatrix}
$$

In the SYCL-MLIR compiler, the \emph{Memory Access Analysis} is currently used by the \emph{Loop Internalization} optimization (\cref{sec:loop-internalization}) to identify SYCL array accesses to consider as candidates for prefetching into local memory.

\section{Device Optimizations}\label{sec:opt-device}
This section describes optimizations designed to improve performance of SYCL device code. These transformations provide a significant speedup for the \texttt{polyhedral} benchmarks in the SYCL-Bench suite~\cite{sycl-bench1} as described in detail in \cref{sec:eval}.

\subsection{Loop Invariant Code Motion (LICM)}\label{sec:licm}
The MLIR community provides a utility which can be used to hoist operations that are free of memory effects out of a region. The SYCL-MLIR compiler implements an LICM transformation which also considers operations that read or store from memory. 
The transformation uses the MLIR memory effect interface that is also used for the reaching definition analysis to determine the memory effects of an operation, and leverages the alias analysis specialized for SYCL (\cref{sec:alias-analysis}) to determine whether memory accesses are aliased. 

An operation can be hoisted if its operands are either already defined outside of the loop, or they can be hoisted out of the loop. Operations with read-only memory effects can be hoisted, as long as the compiler can prove that no operation in the loop may write to the memory locations being read by the candidate operation. Given a candidate SYCL read operation, which loads an operand, the transformation analyzes all values that may alias to it, and determines whether any of them is either defined outside the loop or can also be hoisted. Operations that exhibit a write memory effect can be hoisted if there is no subsequent operation that writes to the same memory location, or an operation that might read from the memory location the candidate operation writes.

Once a loop has been analyzed, and candidate operations for hoisting are identified, the transformation guards the loop by injecting a versioning condition in order to guarantee that the loop is executed at least once (otherwise a hoisted operation might cause a side effect that would not exist in the original code).

The transformation has also the ability to collect candidate operations that cannot be hoisted unless some of their operands are proven at runtime to not alias. These candidates, if they exist, are handled by versioning the transformed loop with a versioning condition to check that the operands preventing hoisting do not overlap in memory.

\subsection{Detect Reduction}\label{sec:array-reduction}

\begin{listing}[t]
\begin{mlir}
affine.for 
  affine.store 
}
\end{mlir}
\caption{Reduction example.}\label{lst:reduction-example}
\end{listing}

\begin{listing}
\begin{mlir}
                       iter_args(
  affine.yield 
}
affine.store 
\end{mlir}
\caption{Transformed version of the reduction in \cref{lst:reduction-example}.}\label{lst:reduction-optimized}
\end{listing}

Array reductions are a common operation in scientific code. This pass is designed to detect this pattern and accumulate the reduction in a scalar variable rather than updating the array element at every loop iteration. As an example, consider the code in \cref{lst:reduction-example}. The loop loads a value from memory at line 2 and updates it at line 5. If the loop iterates $N$ times, it will perform $2N$ memory accesses. Given that \texttt{\%ptr} is loop invariant, this code can be transformed into the code shown in \cref{lst:reduction-optimized}.

The transformed loop receives the incoming value for the array element (line 1) through the \texttt{iter\_args} operand of the \texttt{affine.for} operation. It then accumulates the reduction results in the loop-carried scalar variable \texttt{\%red}. Finally, the array element is updated with the reduction result at line 8. The optimized version of the loop no longer performs any memory accesses involving the array element pointed to by \texttt{\%ptr}. Reducing memory traffic to and from memory is an important optimization on many types of devices and its benefit will be quantified in \cref{sec:eval}.

Alias analysis (\cref{sec:alias-analysis}) is used to ensure the safety of the transformation because in the example above, \texttt{\%ptr} and \texttt{\%other\_ptr} must not be aliased in order for this transformation to be legal.

\subsection{Loop Internalization}\label{sec:loop-internalization}
As discussed in \cref{sec:sycl}, the SYCL programming model defines a memory hierarchy with different types of memory. The different memories in the hierarchy each have unique characteristics regarding size and access latency, opening the possibility for optimization. 

For example, as local memory is smaller but faster than global memory, it can be profitably used to prefetch subsets of a memory region, especially if its accesses exhibit temporal locality or an access pattern that is not conducive to being coalesced by the GPU hardware. 

Using the \emph{Loop Internalization} pass, SYCL memory accesses in perfectly nested loops are made to use local memory by leveraging the MLIR's loop tiling infrastructure. The pass leverages the \emph{Memory Access Analysis} (\cref{sec:memory-access-analysis}) to determine the memory access pattern inside a loop.

The access pattern information from the analysis is then used 
to determine whether the memory access can be coalesced. To do so, the submatrix describing the \emph{inter--work-item} access pattern is obtained by removing columns corresponding to loop induction variables. 

In the example in \cref{sec:memory-access-analysis}, this submatrix is given by the first two columns because the rightmost column corresponds to the loop induction variable.

The access can be coalesced if the inter--work-item access matrix is \emph{Linear} or \emph{ReverseLinear} (as described in~\cite{mem-access-pattern}). Temporal reuse is present if the \emph{intra--work-item} memory access matrix (obtained by removing columns corresponding to thread variables) is not the zero matrix. 
Based on this information, the memory access is assigned to either remain in global memory or to be prefetched into local memory.

\begin{listing}[t]
\begin{sycl}
range<2> global_size(N, N), wg_size(M, M);
cgh.parallel_for<matrix_multiply>(
nd_range<2>(global_size, wg_size), [=](nd_item<2> item) {
  size_t i = item.get_global_id(0),
         j = item.get_global_id(1);
  for (size_t k = 0; k < N; k++)
    C[i][j] += A[i][k] * B[k][j];
});
\end{sycl}
\caption{Command-group function snippet with a loop internalization candidate.}\label{lst:loop-internalization-example}
\end{listing}

\begin{listing}[t]
\begin{sycl}
range<2> global_size(N, N), wg_size(M, M);
local_accessor<float> A_tile(wg_size, cgh);
local_accessor<float> B_tile(wg_size, cgh);
cgh.parallel_for<matrix_multiply>(
  nd_range<2>(global_size, wg_size), 
  [=](nd_item<2> item) {
    size_t i = item.get_global_id(0), 
           j = item.get_global_id(1);
    size_t x = item.get_local_id(0), 
           y = item.get_local_id(1);

    group<2> group = item.get_group();
    for (size_t t = 0; t < N; t += M) {
      A_tile[x][y] = A[i][t + y];
      B_tile[x][y] = B[t + x][j];
      group_barrier(group);
      for (int k = 0; k < M; k++)
        C[i][j] += A_tile[x][k] * B_tile[k][y];
      group_barrier(group);
    }
});
\end{sycl}
\caption{Command-group function snippet from \cref{lst:loop-internalization-example} after loop internalization.}\label{lst:loop-internalization-optimized}
\end{listing}

To demonstrate the transformation performed by the pass, consider the SYCL code in \cref{lst:loop-internalization-example} containing a loop (lines 6--7) within a SYCL kernel launched via \texttt{parallel\_for}.
The load operations corresponding to accessors \texttt{A} and \texttt{B} (line 7) are classified as candidates for using local memory by the analysis, because they both exhibit temporal reuse. When there exists at least one candidate access to be prefetched, the optimization transforms the loop into the MLIR equivalent of the code in \cref{lst:loop-internalization-optimized}. Note that the optimized code requires two barriers, so, before doing the transformation, the \emph{Uniformity Analysis} (\cref{sec:uniformity-analysis}) is used to ensure the loop is not in a divergent region.

In the optimized code, the loop is being tiled by the work-group size $M$ (line 13) and an ${M \times M}$ tile of local memory is allocated for each global memory region (lines 2--3). In the outer loop, a portion of each memory region is prefetched into local memory (lines 14--15). The optimization relies on each work-item in the work-group to initialize the local memory used by the inner loop, and so a group barrier is injected to ensure all threads complete the initialization (line 16). In the tiled inner loop (lines 17--18), the original accesses are substituted by the local accessors (line 18). Finally, a second group barrier (line 19) is injected, to guarantee that the inner loop is completed by all work-items in the work-group before prefetching the next global memory portions. 

The performance gains obtained by this transformation will be discussed in \cref{sec:eval}.

\section{Host raising and host-device optimization}\label{sec:raise-host}

\subsection{Host raising}\label{sec:host-raising}
One of the rationales behind SYCL-MLIR's compilation flow is performing host-code analysis to leverage device code compilation. However, the MLIR code obtained from LLVM IR is too low-level for analysis, as, in the end, there is a one-to-one correspondence between both modules.

In order to obtain a higher-level analysis-friendly representation, we defined an MLIR transformation pass matching patterns present in DPC++'s runtime code, and replacing them with operations in the SYCL dialect. The two main patterns to be raised to perform SYCL-specific host-device optimizations are SYCL objects construction, and kernel scheduling.

\begin{listing}[t]
\begin{sycl}
constexpr std::size_t size = 1024;
...
queue.submit([&](handler &cgh) {
  accessor a(buff_a, cgh, size, read_only);
  accessor b(buff_b, cgh, size, read_only);
  accessor c(buff_c, cgh, size, write_only);
  cgh.parallel_for<K>(size, ...);
});
\end{sycl}
\caption{SYCL CGF example.}\label{lst:raising-source}
\end{listing}

As an example of the raising process, we will use the SYCL program in \cref{lst:raising-source}, which, after compilation, translation to MLIR, and raising, is transformed into the code in \cref{lst:raising-res}. As we can see, the \texttt{sycl.host.*} operations capture all the relevant semantics in the original program according to the aforementioned transformations.

After these higher-level operations capturing SYCL domain-specific semantics have been introduced in the code, static host code analysis can be run to infer relevant properties that will result in device code optimizations. Our analyses make use of the reaching definition analysis described in \cref{sec:reach-def-anal} in order to infer properties such as accessor aliasing or argument constness, including implicit arguments like accessors ranges and offset and the kernel ND-range.

\subsection{Host-device optimization}\label{sec:host-device-optimization}
Our approach for joint analysis of host and device code enables several kinds of optimizations. While at the time of writing, the focus of implementation has been on host-device constant propagation, further optimizations discussed at the end of this section can be implemented in the future.

Exploiting the SYCL dialect domain-specific semantics, two additional kinds of optimizations can be enabled next to conventional constant propagation:

\paragraph*{Constant ND-range propagation} SYCL kernels may use ND-range information (see \cref{sec:sycl}) in their body, e.g., to iterate over a container. These queries are usually implemented as calls to platform-specific built-in functions, which are encoded as SYCL dialect operations in SYCL-MLIR\@. Corresponding getter operations for constant ND-range information are replaced by a constant \texttt{range} or \texttt{id}.

\paragraph*{Accessor members propagation} DPC++ accessors are passed as four kernel arguments (see \cref{lst:raising-res,sec:sycl} for context). Exploiting this argument flattening, and using host code static analysis, we cannot only propagate constant accessor members, but also infer when both ranges are the same, thus replacing uses of one of the argument ranges with the other even when these are not constant.

Note that, thanks to SYCL's heterogeneous nature, constant propagation will result in both host and device code optimizations. On the device side, constant propagation may lead to code optimizations like expressions or control flow simplification. Also, DPC++'s SYCL pipeline includes a \emph{SYCL Dead Argument Elimination pass}, which marks kernel arguments as unused. Using this information, the SYCL runtime will not pass these arguments to the kernel, making kernel launches more efficient on the host side.

\begin{listing}[t]
\begin{mlir}
llvm.function @cgf(
  ...
  sycl.host.constructor(
      {type = !sycl_buffer} : !llvm.ptr, i64
  sycl.host.constructor(
      {type = !sycl_accessor} 
      : !llvm.ptr, !llvm.ptr, !llvm.ptr, !llvm.ptr
  ...
  sycl.host.schedule_kernel 
      [range 
      : !llvm.ptr, !llvm.ptr, !llvm.ptr, 
        !llvm.ptr, !llvm.ptr
}
\end{mlir}
\caption{SYCL CGF in \cref{lst:raising-source} after compilation and raising.}\label{lst:raising-res}
\end{listing}

Performance improvements arising from host-device constant propagation will be discussed in \cref{sec:eval}.

In future work, joint analysis of host and device code can be leveraged for even more analyses and optimizations. One example of that would be to refine alias analysis for SYCL accessors. Many of the device analyses and optimizations in \cref{sec:analyses,sec:opt-device} depend on alias analysis, so providing better alias analysis results for SYCL constructs will therefore enable more powerful analyses and optimizations.

To explain how joint analysis of host and device code can refine alias analysis results, consider the SYCL example in \cref{lst:raising-source} again. An alias analysis considering only device code would neither see the construction of the three accessors (line 4--6) nor the construction of the underlying buffers (omitted for brevity). It would therefore need to assume that the accessors and their underlying pointers may alias, as the SYCL specification allows two accessors to be defined on the same buffer or two different buffers to be overlapping sub-buffers of another buffer.

Joint analysis of the host and device on the other hand would be able to see the construction of the accessors and their underlying buffers in host code and would therefore in many cases be able to determine whether two accessors can alias or not, giving an example of how joint analyses of host and device code can be further extended in the future.

Another class of optimization that would be enabled through joint optimization of host and device affects the application as a whole. One example for this kind of optimization is the fusion of device kernels. By merging multiple SYCL device kernels, the overhead associated with kernel launch can be reduced and dataflow that happens via expensive loads and stores to and from global memory can potentially be made internal to the fused kernel. For SYCL, this was successfully demonstrated by P\'erez et al.\ in~\cite{kernel-fusion}. In their work, because at compilation time host and device were compiled separately, they had to perform fusion at runtime using a JIT compiler, which carries additional overhead that can be only partially mitigated using a compilation cache. With joint analysis and optimization of host and device code, such transformations could be done at compilation time, reducing the runtime overhead.

\section{Evaluation}\label{sec:eval}
\blfootnote{
Notices and Disclaimers: Performance varies by use, configuration and other factors.
Performance results are based on testing on 2023-11-17 and may not reflect all publicly available updates. See \cref{sec:eval} for configuration details.
No product or component can be absolutely secure.
Your cost and results may vary.
Intel technologies may require enabled hardware, software or service activation.
}
To investigate the benefits of the compilation flow described in \cref{sec:flow} and the applicability and effectiveness of the optimizations described in \cref{sec:opt-device,sec:raise-host}, this section presents performance evaluations for typical SYCL applications.

We use the SYCL-Bench benchmark suite~\cite{sycl-bench1,sycl-bench2} as representative examples of SYCL applications. It comprises a number of benchmarks with SYCL kernels often found in scientific applications and HPC workloads in different categories. As the runtime component of the SYCL implementation remains completely unchanged for the SYCL-MLIR compiler and is the same as is used by the baseline compiler, the categories \textbf{runtime} and \textbf{micro}, which mainly test the performance of the runtime component and the GPU device itself, are omitted here. Instead, the focus will be on the benchmarks in the \textbf{polybench} and \textbf{single-kernel} categories. 

The \textbf{polybench} category contains a number of core workloads from domains such as linear algebra found in many HPC applications. Using the defaults provided by the run-script found in the SYCL-Bench suite, a problem size of $1024$ is used for the majority of the benchmarks ({\sf\small 2mm}, {\sf\small 3D Convolution},  {\sf\small 3mm}, {\sf\small Correlation}, {\sf\small Covariance}, {\sf\small FDTD2D}, {\sf\small Gramschmidt}, {\sf\small GEMM}, {\sf\small SYRK} and {\sf\small SYR2K}), while a problem size of $4096$ is used for {\sf\small Atax} and {\sf\small 2D Convolution}, and a size of $16384$ is used for {\sf\small Bicg}, {\sf\small GESUMMV} and {\sf\small MVT}.

The \textbf{single-kernel} category contains real-world applications and kernels from domains such as image processing and molecular dynamics. Again using the defaults from SYCL-Bench, most benchmarks are executed with a problem size of $1,048,576$ ({\sf\small KMeans}, {\sf\small Linear Regression Coeff.}, {\sf\small Molecular Dynamics}, {\sf\small Scalar Product} and {\sf\small Vector Addition}), while {\sf\small Linear Regression} uses a size of $65,536$ and {\sf\small NBody} uses a size of $1024$.

All performance measurements are performed on system with an Intel Xeon Platinum 8480+ CPU with 503 GiB RAM and an Intel Data Center GPU Max 1100 GPU with 48GB RAM, running Ubuntu 22.04.2 LTS (Linux kernel 5.15.0), DPC++ version \texttt{3482e2d}\footnote{Commit \texttt{3482e2d6ecb3adf3c53c0e4a7d2b9d2772c498fe} in https://github.com/intel/llvm/tree/sycl-mlir.} and Intel Level Zero driver version 1.3.26690, and the reported runtimes and speedups are discarding the first run for warm-up of the device driver and then average over thirty runs.

Using this methodology, we compare SYCL-MLIR to two different SYCL implementations, in order to assess the overall performance impact of the compilation flow and optimizations developed in this work.
The first implementation is Intel's state-of-the-art, open-source DPC++ compiler~\cite{dpcpp}. DPC++ is a typical example of an LLVM IR-based SYCL compiler which uses the multi-pass compilation approach discussed in \cref{sec:flow} and shown in \cref{img:combined-comp}. It also shares the SYCL runtime implementation with SYCL-MLIR, making performance better attributable to changes to the compiler itself.

\begin{figure}
    \centering
    \includegraphics[width=\columnwidth]{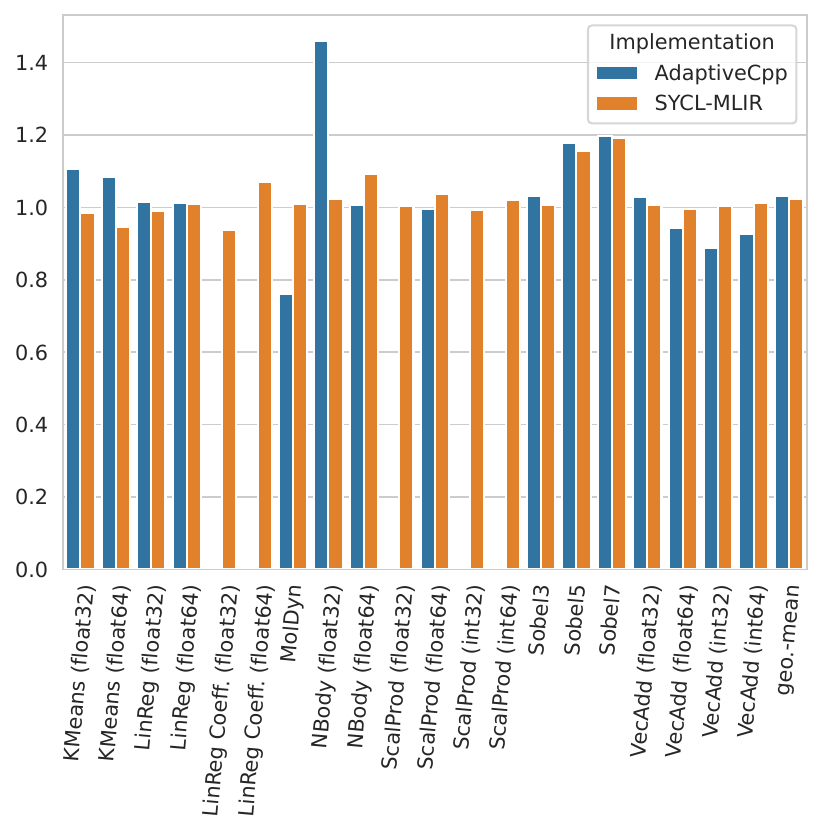}
    \caption{Performance comparison for single kernel benchmarks between DPC++, AdaptiveCpp and SYCL-MLIR with all optimizations enabled. Plot shows speedup over DPC++, higher is better. Missing bars indicate that AdaptiveCpp results failed validation.}
    \label{fig:perf-single-kernel}
\end{figure}

\begin{figure}
    \centering
    \includegraphics[width=\columnwidth]{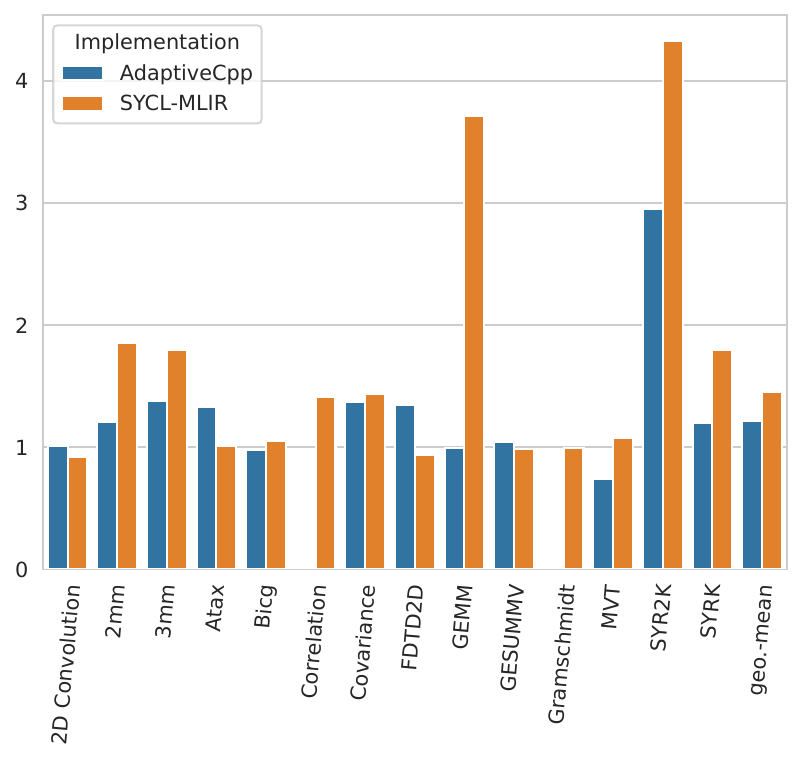}
    \caption{Performance comparison for Polybench benchmarks between DPC++, AdaptiveCpp and SYCL-MLIR with all optimizations enabled. Plot shows speedup over DPC++, higher is better. Missing bars indicate that AdaptiveCpp results failed validation. Note the different y-axis scale compared to \cref{fig:perf-single-kernel}.}
    \label{fig:perf-polybench}
\end{figure}

The second implementation is AdaptiveCpp~\cite{acpp2}. AdaptiveCpp brings its own SYCL runtime implementation and a different compilation flow (cf. \cref{sec:related}), which involves JIT compilation and allows it to leverage runtime information for compilation. We have built AdaptiveCpp version \texttt{c33e8c4}\footnote{Commit \texttt{c33e8c401f46122f2a5b028a46bbecadfb983612} in https://github.com/AdaptiveCpp/AdaptiveCpp.} with LLVM release 17 and Boost version 1.77. The validation of results failed for a number of benchmarks with AdaptiveCpp, indicated by missing bars in \cref{fig:perf-single-kernel,fig:perf-polybench}.

The performance comparison for the benchmarks from the \textbf{single-kernel} category is shown as speedup of SYCL-MLIR and AdaptiveCpp over DPC++ in \cref{fig:perf-single-kernel}. AdaptiveCpp can achieve speedups in a number of cases, but also suffers from a number of mostly smaller degradations, with an overall geo.-mean speedup of 1.03x. The performance benefit of SYCL-MLIR for these benchmarks is small but notable, with small speedups for many benchmarks and a few small performance degradations, yielding a geometric mean speedup of 1.02x.

For the benchmarks from the \textbf{polybench} category, the picture as shown in \cref{fig:perf-polybench} is different. AdaptiveCpp achieves a number of notable speedups over DPC++, with the biggest speedup of close to 3x on the {\sf\small SYR2K} benchmark, while performing on par for the remaining cases with the exception of {\sf\small MVT}, geo.-mean 1.22x.

SYCL-MLIR, apart from a few minor performance regressions, achieves speedups of up to 4.32x over DPC++, with a geo.-mean of 1.45x. 

Multiple benchmarks, including {\sf\small Correlation} and {\sf\small Covariance}, benefit significantly from the array reduction optimization (cf. \cref{sec:array-reduction}). These benchmarks typically contain several opportunities for array reduction, for example {\sf\small Correlation} and {\sf\small Covariance} contain five and four opportunities respectively.

For even more benchmarks, the loop internalization optimization (cf. \cref{sec:loop-internalization}) can improve performance. In particular, five benchmarks ({\sf\small 2mm}, {\sf\small 3mm}, {\sf\small GEMM}, {\sf\small SYR2K}, and {\sf\small SYRK}) benefit significantly from this optimization. Amongst these benchmarks, compiler traces show that the optimization was able to prefetch several array references to local memory. For example, two array references (in the same loop) were prefetched to local memory for the {\sf\small GEMM} benchmark, and four array references were prefetched for the {\sf\small SYR2K} benchmark, also in the same loop. We also note that the {\sf\small Gramschmidt} benchmark contains a candidate loop in a divergent region, and therefore is not optimized by this transformation. Finally, our current implementation does not consider stores as candidates for using local memory, removing this temporary limitation should provide even more opportunities.

The host-device propagation described in \cref{sec:host-device-optimization} also contributes to the performance improvements. For example, in the {\sf\small Sobel7} benchmark, the Sobel filter declared as a constant array can be propagated to the device code to improve performance.

Overall, on SYCL-Bench, SYCL-MLIR achieves a geo.-mean speedup of 1.18x over DPC++ and also performs better than AdaptiveCpp (geo.-mean 1.13x).

As none of the stencil-based workloads from polybench~\cite{polybench} have been ported to SYCL-Bench so far, we complement our investigation of SYCL-Bench with an additional evaluation of some stencil-based workloads. The SYCL version of these workloads stems from the oneAPI samples repository~\cite{oneapi-samples}, from which multiple workloads were extracted.

The one-dimensional heat transfer ({\sf\small 1d\_HeatTransfer}) workload simulates heat transfer in one dimension. In the samples collection, there are two implementations of this workload given, one using the SYCL buffer and accessor model, and another using the SYCL unified shared memory (USM) feature. Both versions of the workload will be included in the evaluation. For the configuration, the recommended setting with 100 points and 1,000 iteration steps was used.

The {\sf\small iso2dfd} sample from the collection uses a two-dimensional stencil to simulate wave propagation in a two-dimensional isotropic medium. As for the heat transfer example, the recommended configuration with a {1,000 $\times$ 1,000} grid and 2,000 iterations is used for this workload.

Lastly, the {\sf\small jacobi} workload demonstrates the use of the Jacobi iteration to solve a linear system of equations. The implementation in the samples collection contains two kernels to execute on the device. One is the actual computation kernel, and the other ({\sf\small prepare\_for\_next\_iteration}) computes the L1-norm and error to prepare for the next iteration. The latter uses SYCL reductions which are currently not yet supported by the compiler presented in this work. Therefore, the workload was adapted such that the preparation for the next iteration happens on the host rather than on the device. However, the main computation kernel still executes on the device.

In sum, as the 1D heat transfer is provided in two versions, this yields four different stencil-based workloads. Similar to above, the performance of SYCL-MLIR and AdaptiveCpp is reported in comparison with the DPC++ compiler.

AdaptiveCpp achieves an 1.5x speedup on  {\sf\small iso2dfd}, but fails to execute the remaining stencil workloads correctly.

For the 1D heat transfer, there is a small performance degradation with the SYCL-MLIR compiler, resulting in 0.86x speedup for the buffer-based implementation and 0.87x speedup for the USM-based implementation.

For the two other stencil workloads, SYCL-MLIR performs on par with the DPC++ compiler, with a 0.99x speedup on {\sf\small iso2dfd} and 1x speedup on {\sf\small jacobi}. A first investigation shows that none of the device optimizations described in \cref{sec:opt-device} is currently applied to those workloads. Improving the SYCL-MLIR compiler by adding new optimizations and extending existing ones for stencil-based and convolution-based workloads is an important goal for future work.

\section{Related Work}\label{sec:related}

Various approaches to compiling C/C++ code into MLIR are currently under development in the LLVM community.
\emph{Polygeist}~\cite{polygeist} supports transforming a subset of the Clang AST directly into a mix of standard MLIR dialects augmented with a small set of custom operations. Affine loops and other structured control-flow constructs from the input program are maintained and exposed to analyses and transformations.
We extended Polygeist to compile SYCL \emph{device} code, however it currently still lacks the capabilities to fully translate \emph{host} code, as virtual functions and exceptions are not yet supported.
Polygeist is also trying to minimize the introduction of custom operations, leading to the relaxation of some MLIR validation rules. While this is fine in the context in which the tool is being used, integration with the rest of the MLIR ecosystem will potentially be more difficult.
\emph{ClangIR}~\cite{cir} is another LLVM incubator project aiming to enable MLIR-based static analysis and code generation from the Clang AST\@. In contrast to Polygeist, the project plans to develop a dialect covering the entire surface of C/C++ and then lower directly to the LLVM dialect, skipping MLIR's built-in higher-level dialects, which eases the correct representation of the source language's semantics.
ClangIR is still in early development, but is a promising candidate to replace the raising of host code in our proposed flow in the future.
\emph{VAST}~\cite{vast} will employ a ``tower of IRs'' to enable program analysis across different levels of abstraction in the compilation of C/C++ code.

Host-device optimization of heterogeneous programming models has been attempted in prior work.
Singer et al.~\cite{syclops} presented \emph{SYCLOps}, a SYCL-specific translator from LLVM-IR to MLIR's standard dialects. While similar to our work in motivation, their tool only raises the \emph{device} code, and does not define a dialect to capture SYCL-specific operations.
Moses et al.~\cite{gpu-transpilation} extended Polygeist to compile CUDA applications to MLIR, representing host and device code simultaneously, and keeping parallel constructs and barriers intact. This enabled various opportunities for code motion across barriers as well as out of parallel regions.
Tian et al.~\cite{openmp-jit-lto} implemented run-time and link-time specializations, such as constant propagation, of OpenMP kernels in LLVM.

Independent from MLIR, the \emph{AdaptiveCpp} project is taking a different approach to SYCL compilation. Initially, the project (formerly known as \emph{hipSYCL})~\cite{acpp} followed the approach of a library-only implementation of SYCL, mapping SYCL device kernels to suitable lower-level programming models such as OpenMP or CUDA and relying on the corresponding compilers, e.g., Nvidia's {\sf\small nvc++}.

In more recent versions, as the one used in \cref{sec:eval}, AdaptiveCpp also added a compiler-based approach to their implementation~\cite{acpp2}, which is however different from other SYCL compilers. In so-called \emph{single source multiple compiler passes} (SMCP~\cite{sycl}) compile flows such as the DPC++ compilation flow depicted in \cref{img:combined-comp}, the source code is processed multiple times, once during host compilation and once for each device compilation, i.e., per target architecture specified by the user. AdaptiveCpp on the other hand follows a so-called \emph{single source single compiler pass} (SSCP~\cite{sycl}) approach, where the source code is only processed once. To this end, the AdaptiveCpp flow uses LLVM IR as an intermediate exchange format and embeds the LLVM IR for the device code in the application binary during the single compilation pass. At runtime, the LLVM IR for the device kernel is then retrieved upon kernel launch and further compiled to the appropriate device-specific format, e.g., SPIR-V for Intel devices or PTX for Nvidia CUDA GPUs.

The compilation from SYCL source code to LLVM IR and later to device format is common to DPC++ and AdaptiveCpp, but the difference is when the second step happens. For DPC++ this happens at compile time, for AdaptiveCpp at application runtime upon kernel launch. 

As such, the AdaptiveCpp compilation flow has one goal in common with SYCL-MLIR: by only performing the second compilation step at runtime, AdaptiveCpp can propagate information about the context of the device invocation from the host to the device compiler, as this information is available at kernel launch when the compilation happens. SYCL-MLIR also tries to propagate this information, but at application compile time by leveraging the MLIR framework to enable joint analyses of host and device code (cf. \cref{img:combined-comp,sec:raise-host}).

The advantage of AdaptiveCpp's approach is that even runtime values can be taken into account for the device compilation, which may yield more information than available to SYCL-MLIR's joint analysis of host and device code at compilation time. On the other hand, the compilation step at runtime causes additional overhead to kernel launches. Even if that overhead can be reduced by caching compilation, that cache is not persisted between distinct application runs, a disadvantage to approaches such as DPC++ or SYCL-MLIR which only need to compile once. 

\section{Conclusion and Outlook}\label{sec:conclusion}
This work presented our practical experience with building an MLIR-based compiler for the SYCL heterogeneous programming model. 
Leveraging MLIR's dialect framework, the SYCL dialect in this work captures key elements of the SYCL API in host and device code on a high level of abstraction, giving the compile flow access to SYCL semantics, such as work-item parallel execution and device memory access. 

Building on top of the dialect, this work implements powerful device optimizations as well as analyses that reason across the border between SYCL host and device code. 

Using these analyses and optimizations, the compile flow in this work achieves speedups of up to 4.3x over a state-of-the-art, LLVM-based SYCL compiler on a collection of SYCL benchmark applications, and proves particularly effective for loop-based workloads.

Despite the speedup achieved in the evaluation, there is still room for improvement: Limitations of the existing C++ frontends for MLIR currently force the compilation flow to perform raising from LLVM IR to MLIR for SYCL host code, as important constructs such as C++ exceptions are not fully supported yet. This currently limits the ability of the compiler to perform optimizations across the host-device border, such as hoisting device code to the host. With the evolution of the C++ frontends discussed in \cref{sec:related}, future work could implement such optimizations and also reason about the overall structure of the SYCL application to perform more advanced optimizations such as device kernel fusion.

\IEEEtriggeratref{14}
\bibliographystyle{IEEEtran}
\bibliography{IEEEabrv,references}

\end{document}